# Extraction of 3D trajectories of mandibular condyles from 2D real-time MRI

Karyna Isaieva[1,*], Justine Leclère[1,2], Guillaume Paillart[1], Guillaume Drouot[3], Jacques Felblinger[1,3], Xavier Dubernard[4] and Pierre-André Vuissoz[1]

[1] IADI, Université de Lorraine, U1254 INSERM, Nancy, F-54000, France;
[2] Oral Medicine Department, University Hospital of Reims, Reims, France;
[3] CHRU-NANCY, INSERM, Université de Lorraine, CIC, Innovation Technologique, F-54000, Nancy, France;
[4] ENT Department, University Hospital of Reims, Reims, France.
* Correspondence: karyna.isaieva@univ-lorraine.fr

**Abstract**

Computing the trajectories of mandibular condyles directly from MRI could provide a comprehensive examination, allowing for the extraction of both anatomical and kinematic details. This study aimed to investigate the feasibility of extracting 3D condylar trajectories from 2D real-time MRI and to assess their precision.

Twenty healthy subjects underwent real-time MRI while opening and closing their jaws. One axial and two sagittal slices were segmented using a U-Net-based algorithm. The centers of mass of the resulting masks were projected onto the coordinate system based on anatomical markers and temporally adjusted using a common projection. The quality of the computed trajectories was evaluated using metrics designed to estimate movement reproducibility, head motion, and slice placement symmetry.

The segmentation of the axial slices demonstrated good-to-excellent quality; however, the segmentation of the sagittal slices required some fine-tuning. The movement reproducibility was acceptable for most cases; nevertheless, head motion displaced the trajectories by 1 mm on average. The difference in the superior-inferior coordinate of the condyles in the closed jaw position was 1.7 mm on average.

Despite limitations in precision, real-time MRI enables the extraction of condylar trajectories with sufficient accuracy for evaluating clinically relevant parameters such as condyle displacement, trajectories aspect, and symmetry.

# INTRODUCTION

Temporomandibular joint disorder (TMD) is a common condition that needs careful diagnosis and treatment planning. Its typical symptoms comprise jaw pain or dysfunction, earache, headache, and facial pain. Understanding the kinematics of mandibular condyles is crucial for TMD management [1,2].

Computerized mechanical axiography [3] has long been the traditional method for tracking the 3D motion of the jaw [4]. However, modern advancements have led to the development of jaw-tracking systems employing ultrasound [5], and infrared or optical cameras [6–11]. Numerous jaw-tracking devices are commercialized [12] and demonstrate excellent theoretical measurement precision [13] and a good correlation with direct measurements [14]. However, they share common drawbacks. Firstly, many require the cutaneous or intra-oral attachment of markers or sensors, potentially causing discomfort and hygiene issues. Although some approaches utilize artificial intelligence algorithms to mitigate this issue [15], this work is still in progress. Secondly, most rely on extra-oral tracking points, resulting in indirect measurements. Finally, in some cases, they still require supplementation with MRI or CT scans for detailed anatomical information.

CT imaging stands as the gold standard for imaging bone tissues, demonstrating its clinical validity [16]. Cone beam CT scanning can be used for tracking 3D condylar trajectories [14,17]; however, the presence of ionizing radiation becomes an ethical issue when a long recording is required. Additionally, hard tissues are typically injured later than soft tissues like the intra-articular disc or capsule [18]. Therefore MRI, which is the gold standard technique for soft tissue imaging [19,20], is frequently prescribed by clinicians. Despite the MRI's long acquisition time complicating dynamic imaging, multiple acquisition and reconstruction strategies allow fast good-quality recordings [21]. Several studies have shown the applicability of the real-time MRI to temporomandibular joint imaging [22–24]. Complementing static high-resolution anatomical MRI with condylar trajectories extracted directly from dynamic MRI would offer a comprehensive, non-invasive, and ionizing radiation-free examination.

The goal of this work was to propose and evaluate the reliability of an automatic algorithm allowing extraction of 3D trajectories of mandibular condyles from 2D real-time MRI. Firstly, we assess the precision of automatic segmentation predictions. Then, we describe and demonstrate the trajectory extraction method and the refinements applied to minimize errors. Finally, we introduce a set of quality metrics aimed at evaluating the accuracy of the extracted trajectories.

# MATERIALS AND METHODS

## Participants and data acquisition

20 healthy volunteers with no known temporomandibular pathologies were recruited. All participants provided written informed consent. The data was acquired under two approved ethical protocols: "METHODO" (ClinicalTrials.gov Identifier: NCT02887053, approval: CPP EST-III, 08.10.01) and its successor "EDEN" (ClinicalTrials.gov Identifier: NCT05218460, approval: CPP SUD-EST IV, 26.07.21). This study was performed in line with the principles of the Declaration of Helsinki.

The images were acquired on a 3T Siemens Prisma with a radial undersampled 2D FLASH sequence [25]. The dynamic MRI was performed in 3 different planes. Two simultaneously imaged oblique sagittal planes passed through the condyles and the canine teeth (TE/TR=1.35/2.5 ms, FoV=100 mm, 168x168, image acquisition time 52.5 ms). One oblique axial slice traversed both condyles and was parallel to the horizontal

Frankfurt plane (TE/TR=1.47/2.34, FoV=192 mm, 136x136, image acquisition time 21.1 ms). The selected real-time sequences were at the end of the imaging protocol and were launched with a difference of 10-15 minutes between them, with the axial sequence being before the sagittal one. For the two first subjects of the set, there was no simultaneous acquisition of both condyles in sagittal planes. The volunteers were asked to slowly open and close the jaw with a periodicity of approximately 6 seconds; however, the actual speed of movement varied among volunteers.

## **Trajectories extraction**

The trajectories were calculated as schematically described in Fig. 1.

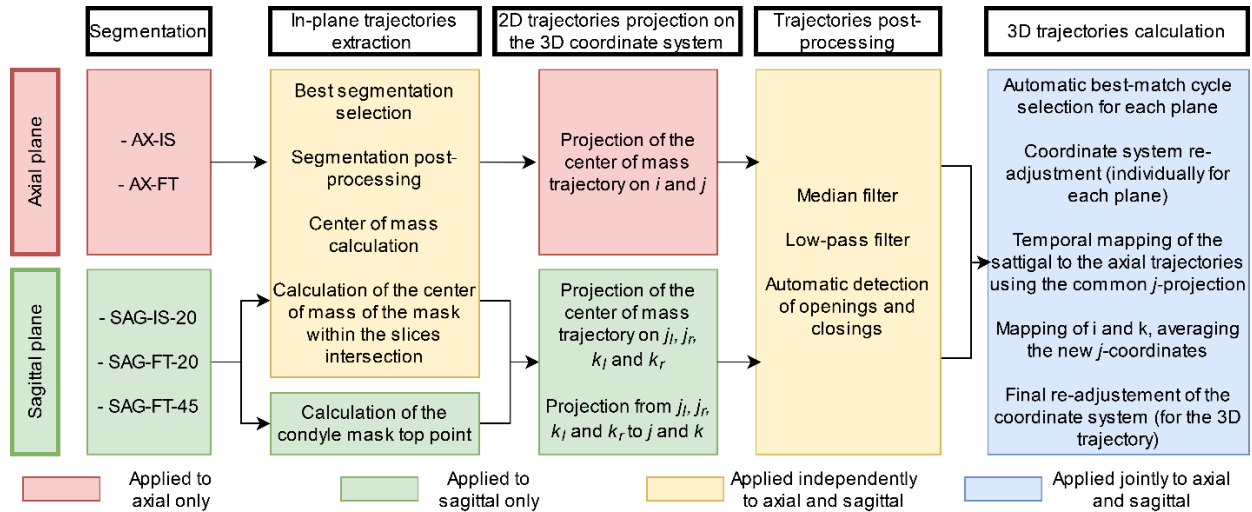

**Fig. 1**. Illustration depicting the process for computing condylar trajectories.

## **Segmentation**

To ensure optimal diversity within the development set, we performed a k-means clustering analysis, similarly to as described in [23]. The selected 20 images were annotated by an expert (JL, a dental surgeon with 10 years of experience). Therefore, a segmented set of 20x20=400 images for each slice orientation and each condyle side was available. The segmentation routine included cropping to a manually selected region of interest (ROI), data augmentation (translations and rotations), and training using a U-Net convolutional neural network [26]. The implementation and training details are described in [23].

To evaluate the impact of this variability on the segmentation results, two different segmentation strategies were explored for the axial plane. The first one (AX-IS, IS for inter-subject) utilized the first 70% of available subjects (14 subjects) as the training set, 10% (2 subjects) as the validation set, and the remaining 20% (4 subjects) as the test set. For the second approach (AX-FT), the model from the IS training was fine-tuned (FT) for each subject. For that, 70% of the images available for each subject (14 images) were used as the training set, 10% (2 images) as the validation set, and 20% (4 images per subject) as the test set. An ROI of 30% of the FOV (58 mm) was selected.

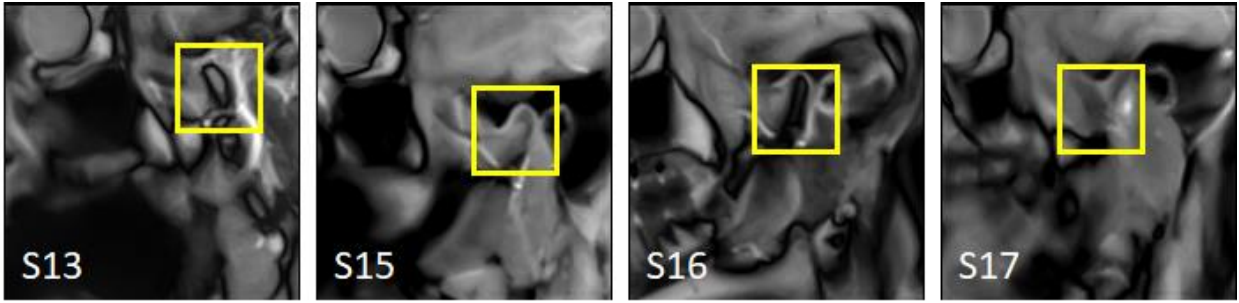

**Fig. 2**. Illustration of the condyle's appearance for different subjects with different placements of the sagittal plane.

The condyles segmentation in the sagittal plane was a challenging task due to the high sensibility of their contrast on the plane position (see Fig. 2). A ROI of 45% of the FOV (45 mm) was sufficient to cover the condyles' motion. However, the preliminary numerical experiments had shown the impossibility of adequately training the IS model using this ROI size. Therefore, three different strategies were tested for the segmentation of the condyles in the sagittal planes. The first one (SAG-IS-20) performed the segmentation within a small ROI of 20% of the initial FOV (20 mm). This ROI size was, however, too small to include the condyles during their motion. Thus, the segmentation was combined with basic center mass tracking. For each frame, a new ROI was centered on the mass center of the previous segmentation mask. The second approach (SAG-FT-20) expanded the first one by fine-tuning the model for each subject. The third one (SAG-FT-45) exploited the IS-trained model to retrain it using the FOV of 45 mm. For the last two strategies, the train-validation-test distribution was the same as before (70%-10%-20%).

The segmentation results were evaluated using two quantitative metrics: Dice similarity coefficient and distance between the center of mass (DCM) of the predicted and the ground truth masks. The results are provided in the format mean± standard deviation. Additionally, an expert performed a blind qualitative evaluation of the whole series' segmentation (JL). For this, video sequences with the segmented condyles were generated in a randomized order. The following scores were used: 0 – Completely failed segmentation, 1 - Bad segmentation quality, serious errors that will substantially change the trajectory, 2 - Medium segmentation quality, overall correct, but imprecise, 3 - Good segmentation quality, there are no visible errors. Additionally, the expert had to select the best segmentation video (only one choice was allowed). The segmentation quality was considered acceptable if it was scored as 2 or 3.

**In-plane trajectories extraction**

The segmentation type was selected for each series based on the expert's decision. The greatest connected component of each segmentation mask was selected (using the MATLAB Image Processing Toolbox) and its center of mass was calculated for each image.

Then, the axial slice was superimposed onto the sagittal images, and vice versa, using positional and orientation data retrieved from the DICOM files. The portions of masks lying inside the slices' intersection volume (referred to as IS-masks) were computed, along with their respective centers of mass.

Additionally, for the sagittal slices, the condyles' top points were calculated. For this, the full segmentation masks were rotated to the principal component space (MATLAB PCA implementation). Only 5% of points with the smallest first principal component were kept and then rotated back into the image space. The average of these points was regarded as the condyle's top point. This approach enhances tracking consistency by focusing on the upper regions of the condyles, as their lower boundaries are not well-defined.

## 2D trajectories projection onto the 3D coordinate system

This step comprised the construction of the 3D coordinate system with the origin $O$ and the orthonormal basis $(\vec{i}, \vec{j}, \vec{k})$ (as shown in Fig. 3a) and projection of the 2D trajectories to the selected coordinate system.

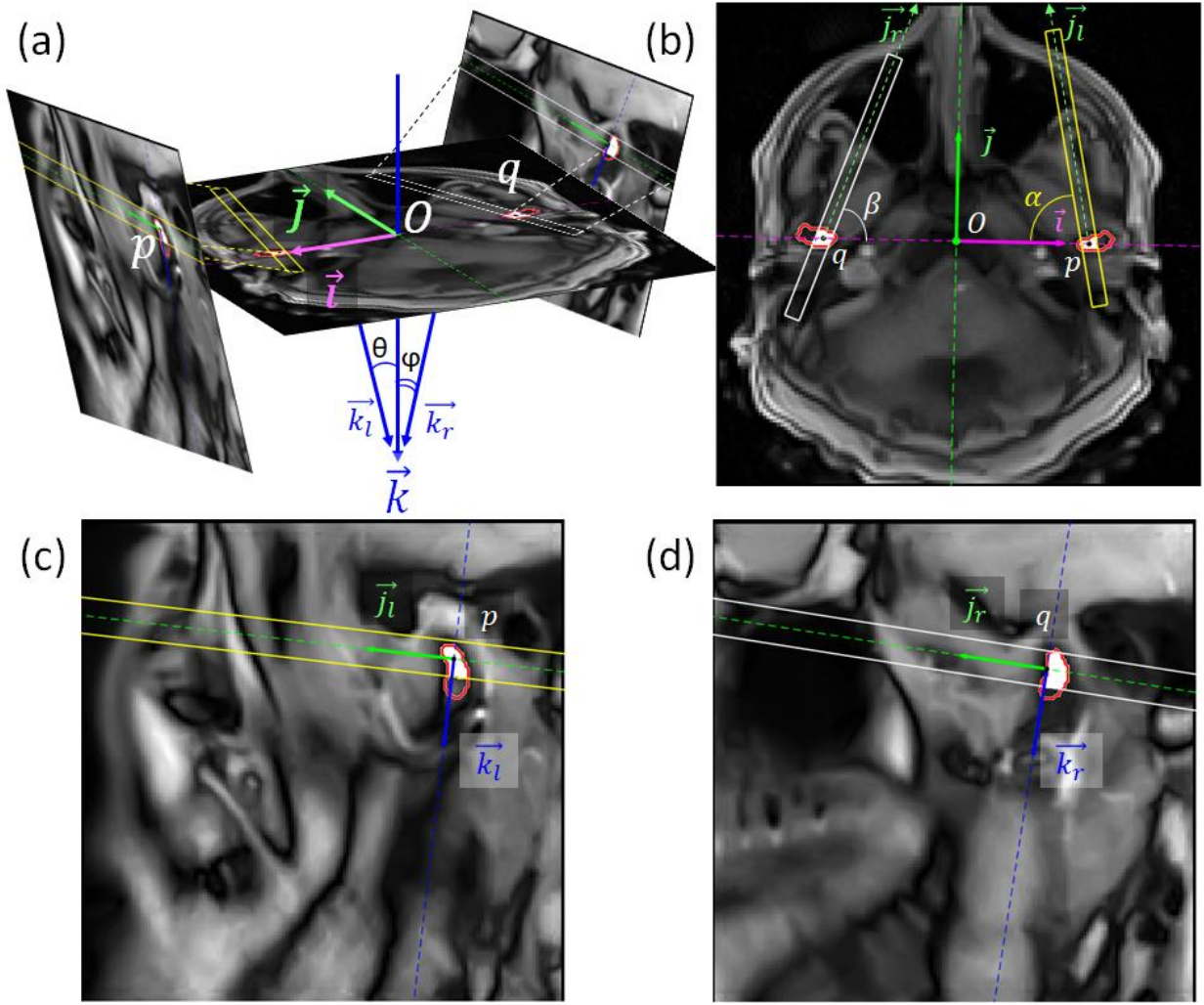

**Fig. 3**. Illustration of the mutual slice orientation and the coordinate system choice. The arrows $\vec{i}$, $\vec{j}$, $\vec{k}$ denote the selected orthonormal basis, $O$ is the origin, and the points $p$ and $q$ show the intersection of the axis $\vec{i}$ with the sagittal slices. The red contours show the predicated condyles' masks and the white area inside them corresponds to the IS-masks. (a) The global 3D view. (b) The axial slice. Yellow rectangle – left sagittal slice, white rectangle – right sagittal slice. (c) Left sagittal slice. Yellow lines – axial slice. (d) Right sagittal slice. White lines – axial slice.

The origin $O$ was defined as the center of mass of the left and right IS-masks from the axial images in the "closed jaw" position. The unitary vector $\vec{i}$ was selected to be parallel to the line traversing these mass centers from right to left (see Fig. 3b). The unitary vector $\vec{j}$ was chosen to lie in the oblique axial plane, be perpendicular to $\vec{i}$ and directed from posterior to anterior. And, finally, the unitary vector $\vec{k}$ was perpendicular to both $\vec{i}$ and $\vec{j}$ and going from superior to inferior.

The 2D trajectories from the axial slice were projected directly to $\vec{i}$ and $\vec{j}$ directions. As for the 2D trajectories derived from the sagittal slice, the origin points—representing the closed jaw position and determined from the axial slice—were identified first: $p$ (for the left sagittal slice) and $q$ (for the right sagittal slice). Following this, the 2D trajectories were projected on directions $\vec{j_l}, \vec{k_l}, \vec{j_r}, \vec{k_r}$, where $j_{l,r}$ is directed parallel to the axial plane and $k_{l,r}$ lies in the sagittal plane, perpendicular to $j_{l,r}$, with $l$ and $r$ denoting the side (left and right) – see Fig. 3b-3d for the illustration. Subsequently, the angles $\alpha$ and $\beta$ (see Fig. 3b) and $\theta$ and $\varphi$ (see Fig. 3a) were calculated. Finally, $\vec{j_l}, \vec{k_l}, \vec{j_r}, \vec{k_r}$ projections were projected to the $(\vec{j}, \vec{k})$ basis.

**Trajectories post-processing**

The trajectories' projections were post-processed with median and adaptive low-pass filtering, as described in [23]. For this, for each of the planes and sides, opening and closing intervals were found using the procedure based on the velocity analysis (described in [23], see Fig. 4a and 4b for an example). A motion cycle was defined as the time interval between two consequent closed jaw positions. Each cycle was associated with a set of three values: averaged IS-mask $j$ (for the left and the right condyle) at the beginning, at the maximum, and at the end of the cycle.

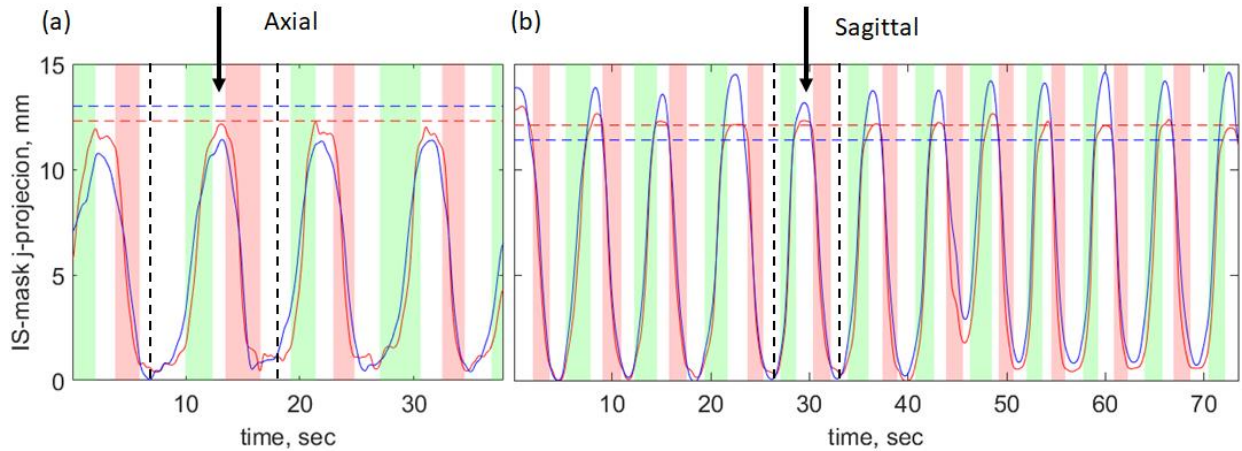

**Fig. 4.** Illustration of the best-match cycles extraction. The red curves correspond to the left condyle and the blue curves correspond to the right condyle. The green and the red areas correspond to the automatically detected openings and closings, correspondently. The black arrows point to the automatically detected best match. The red (for the left condyle) and blue (for the right one) horizontal dashed lines show the mean anterior plateau value calculated from the other slice orientation.

**3D trajectories calculation**

If there was no head motion, and if the opening-closing movements were perfectly reproducible, the j-coordinates of IS-masks from the axial slice should be equal to those from the sagittal slices. In this case, the 3D trajectories computing would be a trivial task. However, human subjects tend to displace their head and do not succeed in exactly reproducing the same movement for each opening and closing cycle. So, the method consisted of attempting to identify the optimal repetition and coordinate systems, and to merge information from both the axial and sagittal slices.

One motion cycle from the axial plane and one from the sagittal plane with the best match of amplitudes were selected. An example of best-match cycles is illustrated in Fig. 4. Afterward, considering possible motion during the acquisition, each slice coordinate system was re-adjusted by translation and rotation in $(i,j)$ plane (axial plane), based on the average between the first and the second closed jaw position as the reference point. All projections extracted previously were translated and rotated to the new coordinate systems.

Then, to temporally match the movement in the axial and the sagittal planes, the IS-masks' $j$-projections were temporally stretched using dynamic time warping (MATLAB Signal Processing Toolbox). The calculated transformation was applied to all projections. The resulting curves were smoothed with MATLAB smoothing splines (with a smoothing parameter of 0.1).

The trajectory $i$-projection was extracted from the axial plane, $k$ – from the sagittal plane, and $j$ was calculated as an average from the axial and the sagittal plane. Finally, the coordinate system was re-adjusted once again, so that both condyles have zero $j$- and $k$ projections at the beginning of the cycle.

## Method's precision evaluation

Despite the minimization of the inconsistencies, some imprecisions were still present. Three possible major sources of these errors are the insufficient reproducibility of the movement, head motion, and imprecise slice placement.

To quantify the first inconsistency, we selected the ratio between the amplitudes of the IS-mask j-projection $A_{AX}/A_{SAG}$ applied to the selected best-match cycles. For the head motion, two metrics were evaluated. The first one assesses the mean-square distance (MSD) between opening and closing. The MSD was defined as follows [27]:

$$MSD(U,V) = \frac{1}{n1+n2}\left(\sum_{i=1}^{n1}\min_j|u_i-v_j| + \sum_{i=1}^{n2}\min_j|u_j-v_i|\right)$$

where $u_i \in U$ and $v_i \in V$ are opening and closing curves, $n1$ and $n2$ is a total number of points in corresponding curves and min refers to the minimization of distance between a point and a curve. The second head motion metrics was selected to be the distance between the initial and the final closed jaw position $d_{init-fin}$. And finally, the slice placement metrics was selected to be the difference between k-projection of the left and the right condyle in the closed jaw position $\Delta k_{L-R}$.

# RESULTS

## Segmentation

The video series with the segmented condyles are presented in [28], and a detailed summary of the quantitative metrics and expert's scores is provided in Online Resource 1. For the axial images, both inter-subject and fine-tuned strategies resulted in good segmentation quality (average score of 2.8 for AX-IS and 2.7 for AX-FT). The most frequent cause of incorrect segmentation was the partial volume effect from the mandibular ramus (as discussed in [23]). For the sagittal slices, the best model was SAG-FT-20 (with an average score of 2.6), followed by SAG-FT-45 (score 2.48). SAG-IS-20 demonstrated the worst results due to tracking errors (score 2.2).

Among the selected segmentations, 18 videos from the axial plane were evaluated with a score of 3, and two videos received a score of 2. For the left sagittal plane, 19 videos scored 3, and one scored 2. For the right sagittal plane, 17 videos received a score of 3, and three received a score of 2. The Dice score for the selected segmentations was 0.84±0.07 for the axial slices and 0.89±0.04 for the sagittal slices, and the DCM was 0.65±0.37 for the axial slices and 1.09±0.70 for the sagittal slices.

## **Trajectories**

The plots of the automatically calculated motion phases are presented in Online Resource 2 and all computed trajectories are provided in Online Resource 3. It should be noted that some subjects were excluded at different stages of the analysis for the reasons given in Table 1.

| Subject | Side | $A_{AX}/A_{SAG}$ | $MSD(op., cl.)$, mm | $d_{init-fin}$, mm | $\Delta k_{L-R}$, mm |
|---|---|---|---|---|---|
| 1 | L | No simultaneous sagittal planes imaging | | | |
|   | R | | | | |
| 2 | L | No simultaneous sagittal planes imaging | | | |
|   | R | | | | |
| 3 | L | 0.96 | 0.28 | 0.29 | 1.29 |
|   | R | 1.02 | 0.29 | 0.44 | |
| 4 | L | 0.97 | 0.39 | 1.25 | 4.49 |
|   | R | 0.92 | 0.39 | 1.18 | |
| 5 | L | 0.97 | 0.89 | 2.54 | -0.37 |
|   | R | 0.9 | 0.72 | 1.93 | |
| 6 | L | No full opening-closing cycle | | | |
|   | R | | | | |
| 7 | L | 0.92 | 0.51 | 1.76 | -0.86 |
|   | R | 0.91 | 0.81 | 2.69 | |
| 8 | L | 0.92 | 0.32 | 1.05 | -1.18 |
|   | R | 0.69 | 0.29 | 1.22 | |
| 9 | L | Masks in the sagittal plane are out of the axial plane | | | |
|   | R | | | | |
| 10 | L | 2.93 | 0.55 | 1.15 | -1.65 |
|    | R | 3.2 | 0.54 | 1.18 | |
| 11 | L | The right condyle in the sagittal plane is out of the axial plane | | | |
|    | R | | | | |
| 12 | L | 1.16 | 0.52 | 0.55 | -2.3 |
|    | R | 1 | 0.8 | 0.6 | |
| 13 | L | Condyles in the sagittal plane are out of the axial plane | | | |
|    | R | | | | |
| 14 | L | 0.92 | 0.52 | 1.96 | 0.32 |
|    | R | 0.86 | 0.73 | 2.54 | |
| 15 | L | The right condyle in the sagittal plane is out of the axial plane | | | |
|    | R | | | | |
| 16 | L | 1.05 | 0.34 | 0.43 | -0.95 |
|    | R | 0.8 | 0.2 | 0.39 | |
| 17 | L | 0.84 | 0.47 | 0.73 | -3.43 |
|    | R | 1.13 | 0.43 | 0.57 | |
| 18 | L | 1.11 | 0.42 | 0.35 | -2.4 |
|    | R | 0.93 | 0.27 | 0.14 | |
| 19 | L | 1.07 | 0.58 | 0.95 | -1.22 |
|    | R | 0.88 | 0.23 | 0.85 | |
| 20 | L | 1.03 | 0.35 | 0.42 | -1.47 |
|    | R | 1.01 | 0.32 | 0.26 | |

**Table 1.** Quality metrics calculated for each extracted trajectory.

Examples of the trajectories extracted from real-time MRI are provided in Fig. 5 and 6.

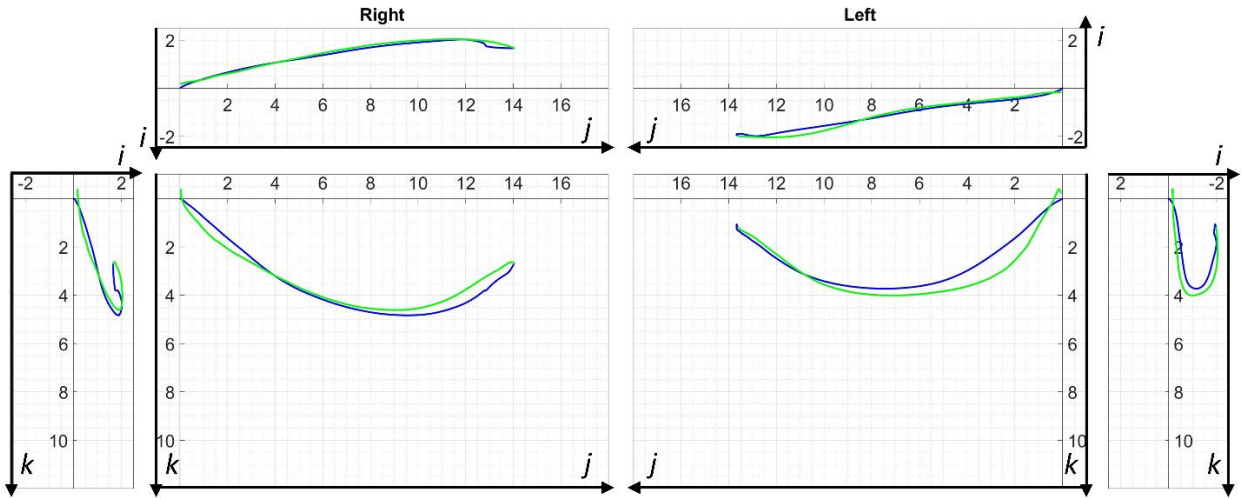

**Fig. 5**. A typical example of extracted mandibular condyles' 3D trajectories during opening-closing motion (subject 3). The blue curves correspond to openings and the green curves – to closings.

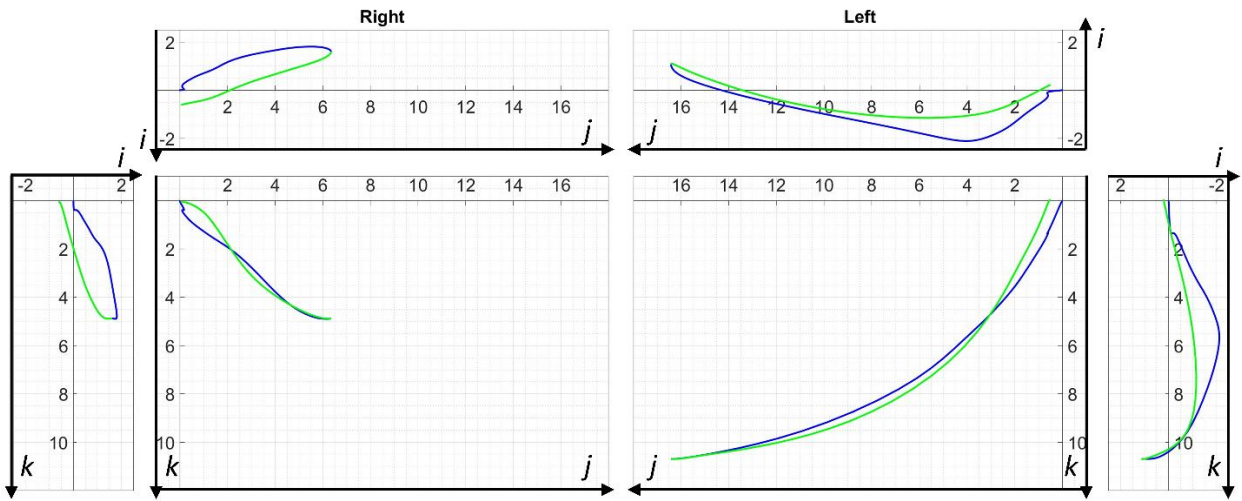

**Fig. 6**. An example of asymmetric mandibular condyles' 3D trajectories during opening-closing motion (subject 12). The blue curves correspond to openings and the green curves – to closings.

The quality metrics of the trajectories are provided in Table 1. The ratio between the $j$-projection amplitudes $A_{AX}/A_{SAG}$ ranged from 0.69 to 3.20 (1.12 on average). An example of IS-mask j-projections illustrating inter-cycle amplitude variation, the best-match cycles, and their amplitudes is presented in Fig. 4.

The MSD between opening and closing path varied from 0.2 mm to 0.89 mm (0.47 mm on average), and the distance between the initial and the final closed jaw coordinate varied from 0.14 mm to 2.69 mm (1.05 mm on average). For some subjects, a systematic displacement of both trajectories in one direction was observed (see Fig. 7); however, for other volunteers, the motion was less predictable.

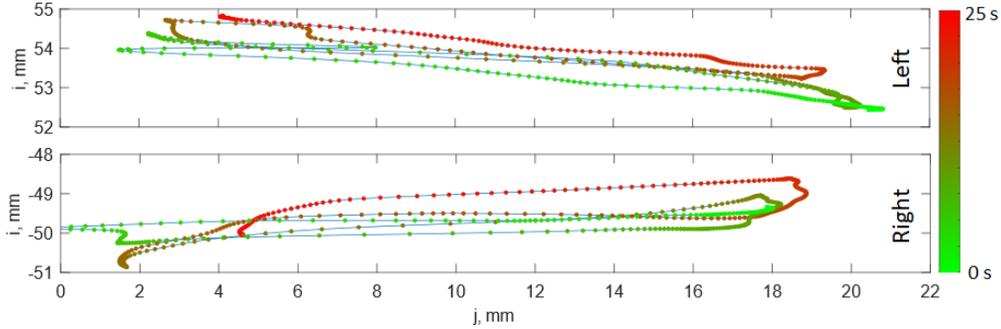

**Fig. 7.** Example of 2D trajectories extracted from the axial plane (subject 4). The color corresponds to the time from the beginning of the sequence. A systematic displacement from right to left can be observed.

The difference in the k-projection in the closed jaw position at the beginning of the selected cycles ranged from -3.43 mm to 4.49 mm (-0.75 mm on average, with the mean of the absolute values of 1.69 mm). An example of asymmetrically places axial slice is provided in Fig. 8.

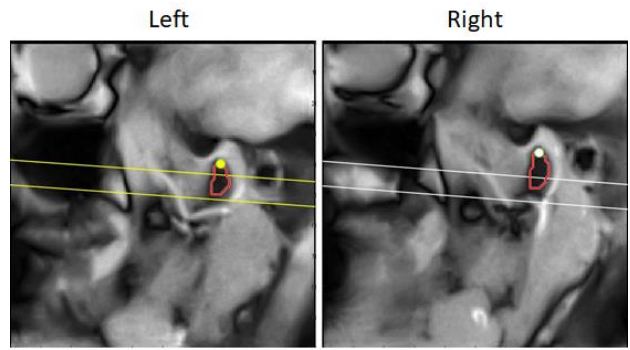

**Fig. 8.** Illustration of an asymmetric axial slice placement. The yellow and the white lines depict the axial slice position in the left and the right sagittal slices, correspondingly, the red curves are the segmentation masks, and the yellow and the white markers point to the top condyle point. The closed jaw position is displayed; however, the right condyle is substantially higher to the axial slice than the left condyle.

## DISCUSSION

We demonstrated that it is possible to compute 3D condylar trajectories from two 2D real-time MRI sequences, one being placed in the oblique axial plane, and one covering 2 oblique sagittal planes. It was shown that the condyles in the axial plane can be automatically segmented with a good segmentation quality. The sagittal images, however, required some fine-tuning. This issue is related to the considerably different contrast of the condyles, as illustrated in Fig. 2. A greater training set should improve the quality of the automatic segmentation in the sagittal plane. Therefore, the proposed trajectories extraction algorithm was almost automated, with the realistic potential to be fully automatic. The appearance of the trajectories is in agreement with the literature [4,9,10].

However, the method has multiple limitations. Firstly, the interslice mapping was based on the j-projection. However, if condyles move mostly out of the Frankfurt plane (like for subject 11), the method is not applicable. In this case, the orientation of the axial plane should be changed. Additionally, more attention should be paid to the placement of the axial plane, which was visibly asymmetric to the anatomical structures for greater absolute values of $\Delta k_{L-R}$. In this case, the trajectories may be artificially asymmetric, while the actual movement was symmetric.

Another serious challenge is related to head motion. While axiographs move along with the head, MRI remains static. Head motion led to the exclusion of subjects 9, 13, and 15 since, during the acquisition of the sagittal slice the condyles were not within the axial slice anymore. This also distorted the extracted trajectories, reflected in high MSD between the opening the closing paths and the distance between the initial and final position. However, both parameters might be high due to other reasons (e.g., the presence of undiagnosed TMD or uncompleted closing). To minimize motion, additional head fixation could be applied. Moreover, in contrast to the protocol utilized in this study, the real-time sequences could be run immediately one after the other, and at the beginning of the imaging protocol. This could also improve the reproducibility of the opening-closing movement which demonstrated an average difference of 12% (approximately 1 mm).

Nevertheless, the computed trajectories could provide a clinical value. Normal condylar trajectories are described in the literature as clear, regular, superimposed curves with a displacement of ≥14 mm [29,30]. Thus, the current precision of the method (which could be further improved considering the proposed protocol changes) still allows the evaluation of their important characteristics, such as length, shape, and symmetry. While small asymmetry of the trajectories can be related to the axial slice placement, the cases of clinically relevant asymmetry, like that demonstrated in Fig. 6, are evident. Similarly, substantially different opening and closing paths may be distinguishable within the obtained precision.

Therefore, despite the method's accuracy being lower than that of state-of-the-art devices [10], MRI-based trajectory computing could be prescribed to patients requiring an MRI exam, as an addition to the conventional protocol.

# CONCLUSIONS

This study is the first to extract 3D trajectories of mandibular condyles, which are relevant clinical information, directly from MRI. We demonstrated the feasibility of their automatic computing from 2D real-time MRI. While there are factors that can compromise the method's precision—such as head motion, asymmetric slice placement, and imperfect movement reproducibility— these influences are manageable and the method still allows for meaningful analysis. We demonstrated that, despite an average shift of approximately 1 mm between two consecutive closed jaw positions, the opening and the closing trajectories were nearly superimposed in the examined healthy subjects (with a mean squared distance of 0.47 mm). Therefore, this method could serve as a complementary tool for patients undergoing MRI for the investigation of the temporomandibular joint soft tissue anatomy, enabling an all-in-one examination.

# ACKNOWLEDGMENT.

This work was supported by the CIC-IT of Nancy. The authors thank Freddy Odille for his useful suggestions. This work was co-funded by the French State-Region contract CPER 2015-2020 (IT2MP), by the European Union through the European Regional Development Fund "FEDER-FSE Lorraine et Massif des Vosges 2014–2020", and France Life Imaging network (grant ANR-11-INBS-0006). The sponsor of this study was the CHRU de Nancy (Department of Methodology, Promotion, Investigation—MPI).

# REFERENCES

[1] Bennett NG. A Contribution to the Study of the Movements of the Mandible. Proceedings of the Royal Society of Medicine 1908;1:79–98. https://doi.org/10.1177/003591570800100813.

[2] Woodford SC, Robinson DL, Mehl A, Lee PVS, Ackland DC. Measurement of normal and pathological mandibular and temporomandibular joint kinematics: A systematic review. Journal of Biomechanics 2020;111:109994. https://doi.org/10.1016/j.jbiomech.2020.109994.

[3] Piehslinger E, Čelar AG, Čelar RM, Slavicek R. Computerized Axiography: Principles and Methods. CRANIO® 1991;9:344–55. https://doi.org/10.1080/08869634.1991.11678382.

[4] Wagner A, Seemann R, Schicho K, Ewers R, Piehslinger E. A comparative analysis of optical and conventional axiography for the analysis of temporomandibular joint movements. The Journal of Prosthetic Dentistry 2003;90:503–9. https://doi.org/10.1016/S0022-3913(03)00482-7.

[5] Sójka A, Huber J, Kaczmarek E, Hędzelek W. Evaluation of Mandibular Movement Functions Using Instrumental Ultrasound System. Journal of Prosthodontics 2017;26:123–8. https://doi.org/10.1111/jopr.12389.

[6] Furtado DA, Pereira AA, Andrade A de O, Bellomo DP, da Silva MR. A specialized motion capture system for real-time analysis of mandibular movements using infrared cameras. BioMedical Engineering OnLine 2013;12:17. https://doi.org/10.1186/1475-925X-12-17.

[7] Pinheiro A, Pereira A, Andrade A, Bellomo Jr D. Measurement of jaw motion: The proposal of a simple and accurate method. Journal of Medical Engineering & Technology 2011;35:125–33. https://doi.org/10.3109/03091902.2010.542270.

[8] Tian S. Three-dimensional mandibular motion trajectory-tracking system based on BP neural network. Mathematical Biosciences and Engineering n.d.;17.

[9] Peck CC, Murray GM, Johnson CWL, Klineberg IJ. The variability of condylar point pathways in open-close jaw movements. The Journal of Prosthetic Dentistry 1997;77:394–404. https://doi.org/10.1016/S0022-3913(97)70165-3.

[10] Bapelle M, Dubromez J, Savoldelli C, Tillier Y, Ehrmann E. Modjaw® device: Analysis of mandibular kinematics recorded for a group of asymptomatic subjects. CRANIO® 2021;0:1–7. https://doi.org/10.1080/08869634.2021.2000790.

[11] Shu J, Ma H, Xiong X, Shao B, Zheng T, Liu Y, et al. Mathematical analysis of the condylar trajectories in asymptomatic subjects during mandibular motions. Med Biol Eng Comput 2021;59:901–11. https://doi.org/10.1007/s11517-021-02346-6.

[12] Revilla-León M, Kois DE, Zeitler JM, Att W, Kois JC. An overview of the digital occlusion technologies: Intraoral scanners, jaw tracking systems, and computerized occlusal analysis devices. Journal of Esthetic and Restorative Dentistry 2023;35:735–44. https://doi.org/10.1111/jerd.13044.

[13] Nagy Z, Mikolicz A, Vag J. *In-vitro* accuracy of a novel jaw-tracking technology. Journal of Dentistry 2023;138:104730. https://doi.org/10.1016/j.jdent.2023.104730.

[14] Lassmann Ł, Nowak Z, Orthlieb J-D, Żółtowska A. Complicated Relationships between Anterior and Condylar Guidance and Their Clinical Implications—Comparison by Cone Beam Computed Tomography and Electronic Axiography—An Observational Cohort Cross-Sectional Study. Life 2023;13:335. https://doi.org/10.3390/life13020335.

[15] Zoss G, Beeler T, Gross M, Bradley D. Accurate markerless jaw tracking for facial performance capture. ACM Trans Graph 2019;38:50:1-50:8. https://doi.org/10.1145/3306346.3323044.

[16] Hilgenberg-Sydney PB, Bonotto DV, Stechman-Neto J, Zwir LF, Pachêco-Pereira C, Canto GDL, et al. Diagnostic validity of CT to assess degenerative temporomandibular joint disease: a systematic review. Dentomaxillofacial Radiology 2018;47:20170389. https://doi.org/10.1259/dmfr.20170389.

[17] Huang C, Xu XL, Sun YC, Guo CB. [A preliminary study on the three-dimensional trajectory of condyle]. Zhonghua Kou Qiang Yi Xue Za Zhi 2018;53:669–73. https://doi.org/10.3760/cma.j.issn.1002-0098.2018.10.005.

[18] Sindelar BJ, Herring SW. Soft Tissue Mechanics of the Temporomandibular Joint. Cells Tissues Organs 2005;180:36–43. https://doi.org/10.1159/000086197.

[19] Ferreira LA, Grossmann E, Januzzi E, Paula MVQ de, Carvalho ACP. Diagnosis of temporomandibular joint disorders: indication of imaging exams. Braz j Otorhinolaryngol 2016;82:341–52. https://doi.org/10.1016/j.bjorl.2015.06.010.

[20] Murphy MK, MacBarb RF, Wong ME, Athanasiou KA. Temporomandibular Joint Disorders: A Review of Etiology, Clinical Management, and Tissue Engineering Strategies. Int J Oral Maxillofac Implants 2013;28:e393–414.

[21] Nayak KS, Lim Y, Campbell-Washburn AE, Steeden J. Real-Time Magnetic Resonance Imaging. Journal of Magnetic Resonance Imaging 2022;55:81–99. https://doi.org/10.1002/jmri.27411.

[22] Krohn S, Gersdorff N, Wassmann T, Merboldt K-D, Joseph AA, Buergers R, et al. Real-time MRI of the temporomandibular joint at 15 frames per second—A feasibility study. European Journal of Radiology 2016;85:2225–30. https://doi.org/10.1016/j.ejrad.2016.10.020.

[23] Isaieva K, Leclère J, Felblinger J, Gillet R, Dubernard X, Vuissoz P-A. Methodology for quantitative evaluation of mandibular condyles motion symmetricity from real-time MRI in the axial plane. Magnetic Resonance Imaging 2023;102:115–25. https://doi.org/10.1016/j.mri.2023.05.006.

[24] Mouchoux J, Meyer-Marcotty P, Sojka F, Dechent P, Klenke D, Wiechens B, et al. Reliability of landmark identification for analysis of the temporomandibular joint in real-time MRI. Head Face Med 2024;20:10. https://doi.org/10.1186/s13005-024-00411-7.

[25] Uecker M, Zhang S, Voit D, Karaus A, Merboldt K-D, Frahm J. Real-time MRI at a resolution of 20 ms. NMR Biomed 2010;23:986–94. https://doi.org/10.1002/nbm.1585.

[26] Ronneberger O, Fischer P, Brox T. U-Net: Convolutional Networks for Biomedical Image Segmentation 2015. https://doi.org/10.48550/arXiv.1505.04597.

[27] Li M, Kambhamettu C, Stone M. Automatic contour tracking in ultrasound images. Clinical Linguistics & Phonetics 2005;19:545–54. https://doi.org/10.1080/02699200500113616.

[28] Isaieva K, Leclère J, Paillart G, Drouot G, Felblinger J, Dubernard X, et al. Videos of axial and sagittal image series with segmented condyles and mutual slice position. Figshare 2024. https://doi.org/10.6084/m9.figshare.25846300.v1.

[29] Talmaceanu D, Bolog N, Leucuta D, Tig IA, Buduru S. Diagnostic use of computerized axiography in TMJ disc displacements. Experimental and Therapeutic Medicine 2022;23:1–8. https://doi.org/10.3892/etm.2022.11137.

[30] Buduru S, Balhuc S, Ciumasu A, Kui A, Ciobanu C, Almasan O, et al. Temporomandibular dysfunction diagnosis by means of computerized axiography. Med Pharm Rep 2020;93:416–21. https://doi.org/10.15386/mpr-1754.